\documentclass[twocolumn,showpacs,amsmath,amssymb,pra]{revtex4}

\usepackage{graphicx}
\usepackage{dcolumn}
\usepackage{bm}
\usepackage{times}
\usepackage{amsmath}

\newcommand{\bra}[1]{\langle #1|}
\newcommand{\ket}[1]{| #1 \rangle}
\newcommand{\braket}[2]{\langle #1|#2 \rangle}

\newcommand{\expi}[1]{ {\rm e}^{{\rm i} #1}}
\newcommand{\expim}[1]{ {\rm e}^{-{\rm i} #1}}
\newcommand{\cre}[1]{\hat{#1}^\dagger}
\newcommand{\des}[1]{\hat{#1}}

\newcommand{\op}[1]{\hat{#1}}

\begin{document}

\title{Generation of macroscopic superposition states in ring superlattices}

\author{ Andreas Nunnenkamp$^{1}$\footnote{Electronic address: a.nunnenkamp1@physics.ox.ac.uk}, Ana Maria Rey$^{2}$ and Keith Burnett$^{1,3}$}

\affiliation{$^{1}$
Clarendon Laboratory, Department of Physics, University of Oxford,
Parks Road, Oxford, OX1 3PU, UK}
\affiliation{$^{2}$ Institute for Theoretical Atomic, Molecular and
Optical Physics, Cambridge, MA, 02138, USA}
\affiliation{$^{3}$ University of Sheffield, Firth Court, Western Bank, Sheffield S10 2TN, UK}

\date{\today}

\begin{abstract}
Ultracold bosons in rotating ring lattices have previously been shown to form macroscopic superpositions of different quasi-momentum states. We demonstrate that the generation of  such kind of states  using slightly non-uniform ring lattices has several advantages: the energy gap decreases less severely with the number of particles, the sensitivity to detunings from the critical rotation frequency is reduced, and the scheme is not limited to commensurate filling. We show that different quasi-momentum states can be distinguished in time-of-flight absorption imaging and propose to probe correlations via the many-body oscillations induced by a sudden change in the rotation frequency.
\end{abstract}


\maketitle

\section{Introduction}

Macroscopic superposition states, to which we refer in this paper as (Schr\"odinger) cat states, are one class of strongly-correlated states which has attracted much interest since the early days of quantum theory \cite{Schrodinger35}. Apart from foundational questions \cite{Leggett02} they have been shown to enable precision measurements at the Heisenberg limit \cite{Bollinger96} and serve as resources for various quantum information tasks \cite{Leibfried05}. Efficient cat state production is therefore an issue of theoretical and practical importance.

Among theoretical proposals for creating cat states with Bose-Einstein condensates \cite{Cirac98, Sorensen01, Dunningham06a, Dunningham06b} one is to load ultracold bosons into rotating ring lattices \cite{Hallwood06a, Hallwood06b, Rey07}. At a critical rotation frequency the two lowest-lying states are the symmetric and anti-symmetric superpositions of different flow around the loop \cite{Hallwood06a, Rey07}. To observe these strongly-correlated states is however largely impossible as their energy difference decreases exponentially in the number of particles \cite{Rey07}. On the other hand, superpositions of macroscopically distinct flow have been observed in experiments with superconducting quantum interference devices (SQUID) \cite{Friedman00, vanderWal00, Chiorescu03}, i.e.~superconducting rings with the superconductivity deliberately destroyed at one or several places in the loop \cite{Tinkham96}.

In this paper we address the question whether cat state production with ultracold bosons in rotating ring lattices can be improved by varying the coupling due to tunneling along the ring. After introducing the twisted Bose-Hubbard Hamiltonian for this problem, we diagonalize analytically the single-particle Hamiltonian for ring superlattices where the tunneling matrix elements alternate along the ring. We show that in slightly non-uniform rings with weak on-site interactions there is a critical rotation frequency at which, similar to the case of a uniform ring lattice, the ground and first excited states become cat-like superpositions of the two lowest-lying degenerate single-particle modes. We demonstrate that their energy difference scales less severely with the number of particles, the superposition states are less sensitive to detunings away from the critical rotation frequency compared to the uniform case, and cat state production is not limited to commensurate filling of the lattice. Finally, we discuss how one can probe cat-like correlations via many-body oscillations induced by a sudden change in the rotation frequency and show that different quasi-momentum states can be distinguished in time-of-flight absorption images.

\section{The Hamiltonian}

We consider a system of $N$ ultracold bosons with mass $M$ confined in a 1D ring superlattice of an even number of sites $L$ with lattice constant $d$. The ring is rotated about its axis ($z$ axis) with angular velocity $\Omega$. A theoretical proposal how to experimentally achieve such a ring-shaped optical lattice can be found in Ref.~\cite{Amico05}. In the rotating frame of the ring the many-body Hamiltonian is given by \cite{Bhat05, Rey07}
\begin{equation}
\hat {H}=\int d\mathbf{x}
\hat{\Phi}^{\dagger}\left[-\frac{\hbar^2}{2 M}\nabla^2 +
V(\mathbf{x}) +\frac{4 \pi \hbar^2 a}{2M}
\hat{\Phi}^{\dagger}\hat{\Phi}- \Omega \hat{L}_z\right]\hat{\Phi}
\end{equation}
where $a$ is the $s$-wave scattering length, $V(\mathbf{x})$ the lattice potential, $\hat{L}_z$ the angular momentum, and $\mathbf{x}$ the 3D spatial coordinate vector. As the lattice potential $V(\mathbf{x})$ is a 1D ring superlattice in the $x-y$ plane, it confines the motion along the $z$ axis as well as the radial direction in the $x-y$ plane so strongly that only the motion of the atoms along the ring has to be taken into account.

Assuming that the lattice is deep enough to restrict tunneling between nearest-neighbor sites and the band gap is larger than the rotational energy, the bosonic field operator $\hat{\Phi}$ can be expanded in Wannier orbitals confined to the first band $\hat{\Phi}=\sum_j \hat{a}_j W_j'(\mathbf{x})$ with $W_j'(\mathbf{x}) = \exp \left[ \frac{-i M}{\hbar} \int_{\mathbf{x}_j'}^\mathbf{x} \mathbf{A} (\mathbf{x}') \cdot d\mathbf{x}' \right] W_j(\mathbf{x})$. Here $W_j(\mathbf{x})$ are the Wannier orbitals of the stationary lattice centered at the site $j$, $\mathbf{A}(\mathbf{x})=\Omega \hat{z} \times \mathbf{x}$ an effective vector potential and $\hat{a}_j$ the bosonic annihilation operator of a particle at site $j$. In terms of these quantities, the many-body Hamiltonian can be written, up to on-site diagonal terms which we neglect for simplicity, as \cite{Bhat05, Hallwood06a, Rey07}
\begin{equation}
\op{H}
= -\sum_{j=1}^L \left( J_j \expi{\theta} \cre{a}_{j+1} \des{a}_j + H.c. \right) + \frac{U}{2} \sum_{j=1}^L \hat{n}_j(\hat{n}_j-1).
\label{ham}
\end{equation}
Here $\hat{n}_j=\hat{a}_j^{\dagger}\hat{a}_{j}$ is the number operator at site $j$, $\theta$ is the effective phase twist induced by the gauge field, $\theta \equiv \int_{\mathbf{x}_{i}}^\mathbf{x_{i+1}} \mathbf{A}(\mathbf{x}')\cdot d\mathbf{x}' = \frac{M \Omega L d^2}{h}$, $J_j$ is the hopping energy between nearest neighbor sites $j$ and $j+1$: $J_j \equiv \int d\mathbf{x} W_j^{*} \left[-\frac{\hbar^2}{2M} \nabla^2 + V(\mathbf{x})\right] W_{j+1}$, and $U$ the on-site interaction energy: $U\equiv \frac{4 \pi a \hbar^2}{M} \int d\mathbf{x}|W_j|^4$.

\begin{figure}
 \centering
 \includegraphics[width=\columnwidth]{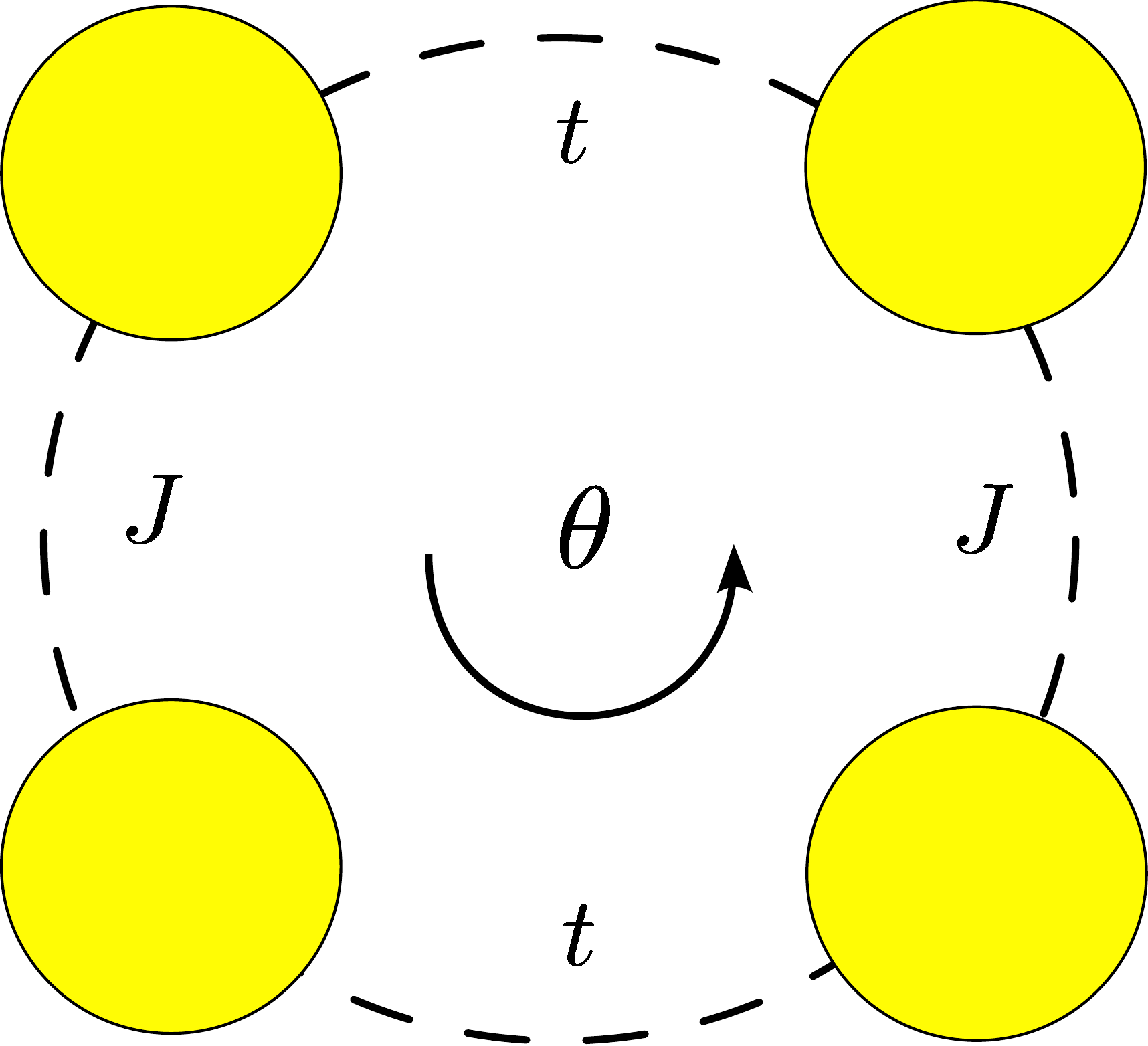}
 \caption{(Color online) One-dimensional ring lattice with $L=4$ sites and alternating tunneling matrix elements $J$ and $t$. In the rotating frame an effective phase twist $\theta$ is induced.}
 \label{fig:setup}
\end{figure}

The case of a uniform ring, i.e.~$J_j = J$ for all $j$, has been discussed before. In this paper we generalize to non-uniform ring lattices where we will focus on superlattice structures where $J_j = J$ for $j$ even and $J_j = t$ for $j$ odd (see Fig.~\ref{fig:setup}). In this case the single-particle problem can be solved analytically.

\section{The non-interacting system}

We shall first study the properties of the Hamiltonian (\ref{ham}) in the limit of vanishing on-site interaction $U = 0$.

Using the decompositions
\begin{equation}
\des{a}_j = \frac{1}{\sqrt{L}} \sum_{q=0}^{L-1} \expi{2\pi q j / L} \des{b}_q
\end{equation}
along with
\begin{equation}
\frac{1}{L} \sum_{j=0}^{L-1} J_j \expim{2\pi (q-q') j / L}
= \left\{ \begin{array}{cl}
- (J+t)/2 & \textrm{for} \, q=q' \\
+ (J-t)/2 & \textrm{for} \, q = q' + L/2 \\
0 & \textrm{otherwise} \\
\end{array} \right.
\end{equation}
we obtain the single-particle Hamiltonian $\op{H}_{\textrm{sp}}$ in the quasi-momentum basis, i.e.
\begin{equation}
\op{H}_{\textrm{sp}}
= \sum_{q=0}^{L/2-1}
\left( \begin{array}{cc} \cre{b}_q & \cre{b}_{q+L/2} \end{array}\right)
\mathbf{M}_q
\left( \begin{array}{c} \des{b}_q \\ \des{b}_{q+L/2} \end{array}\right).
\label{hamsp}
\end{equation}
Here the two-by-two matrices are given by
\begin{equation}
\mathbf{M}_q =
\left( \begin{array}{cc}
- (J+t) \cos \left(\theta-\frac{2\pi q}{L}\right) & - i (J-t) \sin \left(\theta-\frac{2\pi q}{L}\right) \\
+ i (J-t) \sin \left(\theta-\frac{2\pi q}{L}\right) & + (J+t) \cos \left(\theta-\frac{2\pi q}{L}\right)
\end{array} \right)
\end{equation}
and represent the coupling between the quasi-momentum states. We can diagonalize the single-particle Hamiltonian (\ref{hamsp}) via a unitary basis transformation
\begin{equation}
\left( \begin{array}{c} \op{c}_q \\
\op{c}_{q+L/2} \end{array} \right)
= \left( \begin{array}{cc}
\cos \alpha_q & i \sin \alpha_q \\
i \sin \alpha_q & \cos \alpha_q
\end{array} \right)
\left( \begin{array}{c} \op{b}_q \\
\op{b}_{q+L/2} \end{array} \right)
\label{definec}
\end{equation}
with
\begin{equation}
\tan 2\alpha_q = \frac{J-t}{J+t} \cdot \tan\left(\theta - \frac{2\pi q}{L}\right)
\end{equation}
and obtain
\begin{equation}
\op{H}_{\textrm{sp}}
= \sum_{q=0}^{L/2-1}
\left( \begin{array}{cc} \cre{c}_q & \cre{c}_{q+L/2} \end{array}\right)
\left( \begin{array}{cc}
E_{q}^- & 0 \\
0 & E_{q}^+
\end{array} \right)
\left( \begin{array}{c} \des{c}_q \\ \des{c}_{q+L/2} \end{array}\right)
\label{hamspdiag}
\end{equation}
where the single-particle energies are given by
\begin{equation}
E_{q}^{\pm} = \pm \sqrt{J^2 + t^2 + 2J t \cos \left(2\theta-\frac{4\pi q}{L}\right)}
\label{spspectrum}
\end{equation}
which for $t=J$ simplify to \cite{Rey07}
\begin{equation}
E_{q}^{\pm} = \pm 2J \left| \cos \left( \theta-\frac{2\pi q}{L}\right) \right|.
\end{equation}

In the uniform ring, $t/J=1$, the eigenstates of the single-particle Hamiltonian $\op{H}_{\textrm{sp}}$ are quasi-momentum states. At certain phase twists $\theta$ they are twice degenerate. For example, for $L=4$ sites the quasi-momentum states $\ket{q=1}$ and $\ket{q=-1}$ are degenerate at $\theta = 0$, whereas at $\theta=\pi/4$ the states $\ket{q=0}$ and $\ket{q=1}$ as well as $\ket{q=2}$ and $\ket{q=-1}$ are degenerate. Reducing the symmetry of the ring by choosing $t\not=J$ the quasi-momentum states which differ by $L/2$ quasi-momentum units are coupled by the single-particle Hamiltonian (\ref{hamsp}), so that the quasi-momentum states $\ket{q=1}$ and $\ket{q=-1}$ hybridize and the degeneracy at $\theta = 0$ is lifted, i.e.~$E_1^+ - E_1^- = 2(J-t)$.

\begin{figure}
 \centering
 \includegraphics[width=\columnwidth]{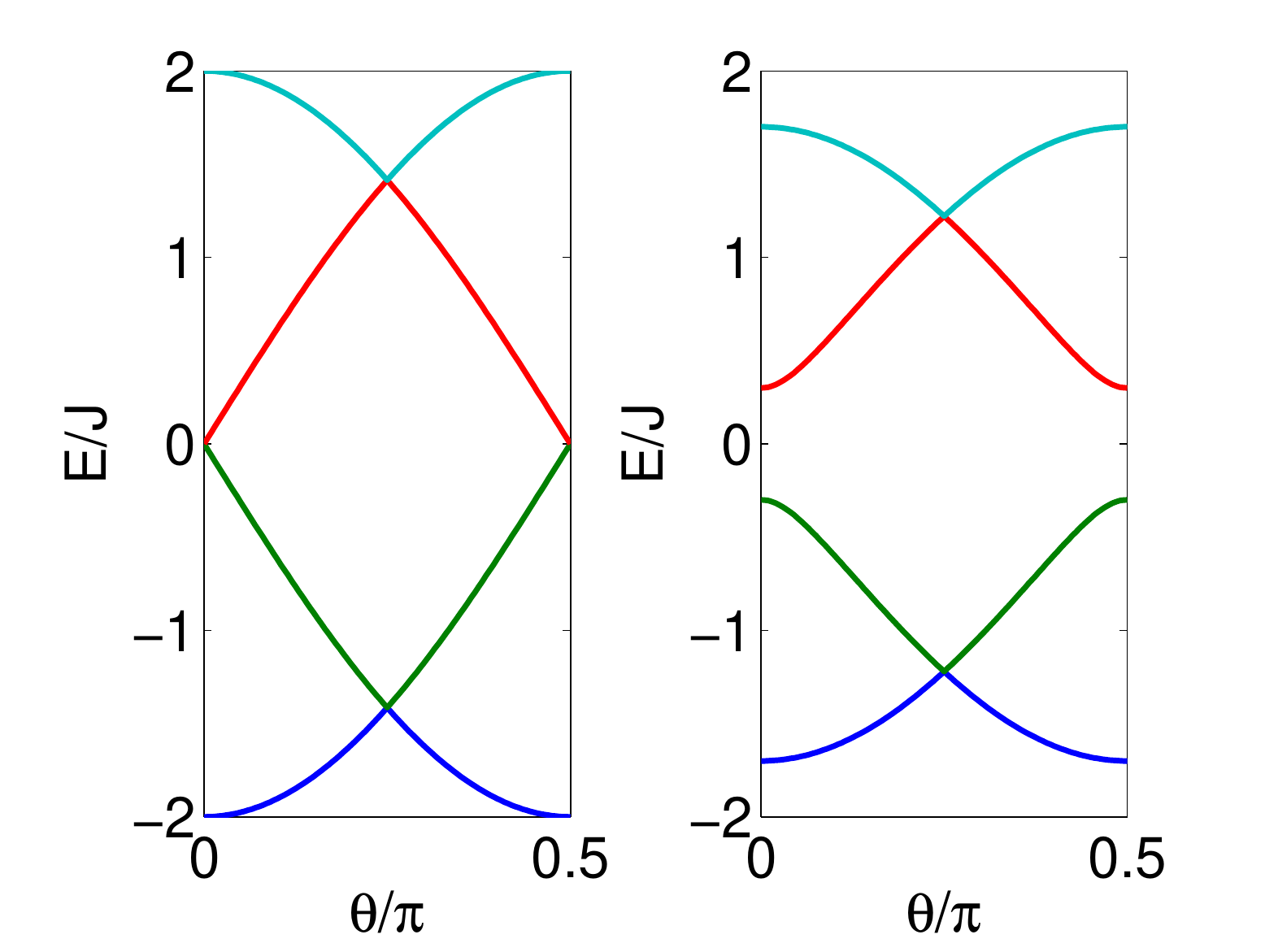}
 \caption{(Color online) Single-particle spectrum for $L=4$ sites with $t/J=1$ (left) and $t/J=0.7$ (right) as a function of the phase twist $\theta$. In both cases there are level crossings at $\theta = \pi/4$.}
 \label{fig:spspectrum}
\end{figure}

At $\theta = \pi/4$ however the degenerate quasi-momentum states are not coupled by the single-particle Hamiltonian (\ref{hamsp}), so that the degeneracy is present also in the non-uniform case. This remains true for arbitrary $L$, i.e.~$E_0^- = E_{L/4}^-$ at $\theta = \pi/4$, but for $L \not= 4$ these states are not the ground states of the system. In Fig.~\ref{fig:spspectrum} we plot the single-particle spectrum for $L=4$ sites as a function of the effective phase twist $\theta$. It shows level crossings at $\theta=\pi/4$ both for $t = J$ as well as $t\neq J$. In the following we will refer to $\theta=\pi/4$ as the critical phase twist, since -- as we will demonstrate below -- weak on-site interactions lift the degeneracy at $\theta=\pi/4$ and lead to the formation of strongly-correlated states in the many-body system.

\section{Effective many-body Hamiltonian}

In this section we derive an effective many-body Hamiltonian to study the properties of the many-body system in the limit of weak on-site interactions. In the absence of interactions $U = 0$ the ground state of a bosonic many-body system is the state with all bosons occupying the lowest-energy single-particle state. At the critical phase twist $\theta = \pi/4$ the single-particle Hamiltonian (\ref{hamsp}) has, however, two degenerate single-particle states, so that there is a $N+1$-dimensional degenerate subspace at $N$-particle level. A convenient basis for this subspace is given by the Fock states $|n,N-n\rangle$ with $0 \le n \le N$, where $n$ particles are in the single-particle state of energy $E_0^{-}$ and $N-n$ particles in the one of energy $E_{L/4}^{-}$, respectively. For weak interactions $NU/L \ll 2\sqrt{J^2+t^2}$ this subspace is the low-energy sector of the many-body problem for all phase twists $\theta$ and tunneling strength ratios $t/J$.

\begin{figure}
 \centering
 \includegraphics[width=\columnwidth]{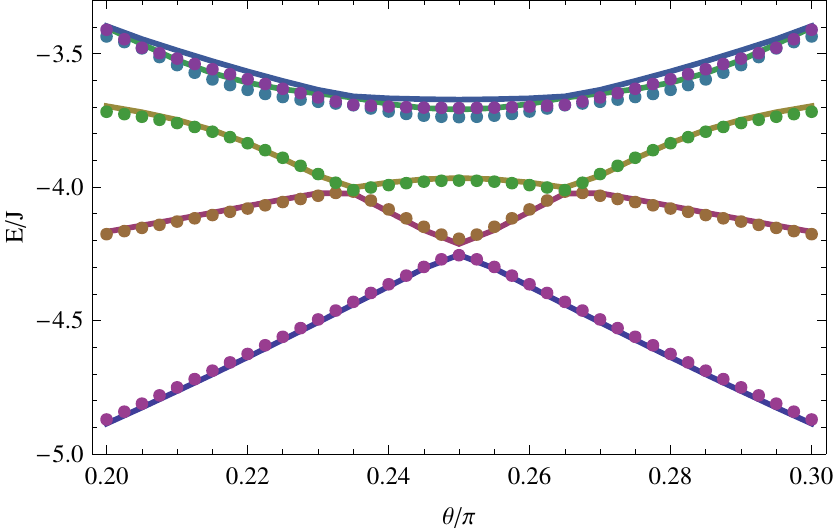}
 \caption{(Color online) Many-body spectrum for $L=N=4$, $t/J=0.7$ and $U/J=0.5$ as a function of the phase twist $\theta$ obtained from exact diagonalization (dotted lines) of the many-body Hamiltonian (\ref{ham}) as well as from the effective Hamiltonian (\ref{heff}) (solid lines).}
 \label{fig:mbspectrum}
\end{figure}

Starting from the interaction Hamiltonian
\begin{eqnarray}
\op{H}_\textrm{int} &=& \frac{U}{2} \sum_{j=1}^L \cre{a}_j \cre{a}_j \des{a}_j \des{a}_j \nonumber \\
&=& \frac{U}{2 L} \sum_{k_1, k_2, k_3 = 0}^{L-1} \cre{b}_{k_1} \cre{b}_{k_2} \des{b}_{k_3} \des{b}_{\left[k_1+k_2-k_3\right] \, \textrm{mod} \, L}
\label{hamint}
\end{eqnarray}
we first restrict the sum in Eq.~(\ref{hamint}) to the relevant modes $k_i \in \{ 0, L/4, L/2, 3L/4 \}$ with $i=\{1,2,3\}$ and second use the inverse of Eq.~(\ref{definec}) to express the operators $\des{b}_q$ in terms of the operators $\des{c}_q$. Keeping only terms within the low-energy subspace we obtain the effective Hamiltonian up to first order in the on-site interaction strength $U$
\begin{eqnarray}
\op{H}_{\textrm{eff}} &=& \left(E_{0}^- \op{n}_0 + E_{L/4}^- \op{n}_{L/4} \right)
+ \frac{U}{2L} \left( 2 \op{n}_0 \op{n}_{L/4} + N^2 - N \right) \nonumber \\
&& + \left( \frac{i \eta U}{2L} \cre{c}_0 \cre{c}_0 \des{c}_{L/4} \des{c}_{L/4} + H.c. \right)
\label{heff}
\end{eqnarray}
where $\op{n}_q = \cre{c}_q \des{c}_q$ are the number operators and the parameter
\begin{eqnarray}
\eta &=& 2 \left(\cos \alpha_0 \sin \alpha_0 \cos 2\alpha_{L/4} \right. \nonumber \\
&& \left. + \cos \alpha_{L/4} (- \sin \alpha_{L/4}) \cos 2\alpha_0 \right)
\end{eqnarray}
satisfies $0 \le \eta \le 1$ and simplifies for $\theta=\pi/4$ to
\begin{equation}
\eta = \frac{J^2-t^2}{J^2+t^2}.
\end{equation}
The first bracket of Eq.~(\ref{heff}) contains the contributions from the single-particle Hamiltonian (\ref{hamspdiag}), whereas the terms in the second and third brackets arise from the on-site interaction (\ref{hamint}). At the critical phase twist $\theta = \pi/4$ the former are an unimportant zero-energy offset, whereas the terms in the second bracket shift the energies of the states in the subspace differently, e.g.~they lead to an energy difference of $U(N-1)/L$ between the states $\ket{N,0}$ and $\ket{N-1,1}$, while the states $\ket{n,N-n}$ and $\ket{N-n,n}$ remain pairwise degenerate. The terms in the third bracket are off-diagonal in the Fock basis of the subspace and describe two-particle scattering between the two single-particle modes. As we will see in the next section they lift the remaining pairwise degeneracies in the many-body spectrum. In Fig.~\ref{fig:mbspectrum} we plot the many-body spectrum for $L=N=4$, $t/J=0.7$ and $U/J=0.5$ as a function of the phase twist $\theta$ obtained from exact diagonalization of (\ref{ham}) and the effective Hamiltonian (\ref{heff}). We see that for these parameters the effective Hamiltonian (\ref{heff}) is an excellent approximation for the low-energy sector of the many-body Hamiltonian (\ref{ham}).

\subsection{Cat-like superpositions in the limit $t/J \approx 1$}

Let us now determine the ground and first excited state for slightly non-uniform rings $t/J \approx 1$, close to the critical phase twist $\theta \approx \pi/4$. Since the terms in the second bracket of Eq.~(\ref{heff}) increase the energy for all states in the subspace apart from $\ket{N,0}$ and $\ket{0,N}$ and the coupling between the states is weak (as the coupling $\eta$ is small in this limit), we project the effective Hamiltonian (\ref{heff}) onto the subspace spanned by these two nearly-degenerate lowest-energy states \cite{Rey07}.

As there is no direct coupling between $\ket{N,0}$ and $\ket{0,N}$ we calculate the total coupling through intermediate states using perturbation theory. After eliminating the intermediate states we obtain the following $2 \times 2$ Hamiltonian
\begin{equation}
\op{H}_{2 \times 2} = \left( \begin{array}{cc}
\Delta E/2 + \xi  & \Delta \\
\Delta^* & - \Delta E/2+ \xi
\end{array}
\right)
\label{ham2by2}
\end{equation}
where $\Delta E$ is the energy difference between the states $\ket{N,0}$ and $\ket{0,N}$ caused by the detuning of the phase twist from resonance $\Delta \theta = \theta - \pi/4$, i.e.
\begin{equation}
\Delta E = N(E_{L/4}^- - E_0^-) \approx \frac{4J t N \Delta \theta}{\sqrt{J^2 + t^2}},
\end{equation}
and $\Delta$ is the coupling between the states $\ket{N,0}$ and $\ket{0,N}$ due to the off-diagonal terms of the effective Hamiltonian (\ref{heff}). As the latter only directly couples the states $|n,N-n\rangle$ and $|n\pm2,N-n\mp 2\rangle$, the first non-vanishing order is given by
\begin{equation}
\Delta = \frac{\bra{N,0} \op{H}_{\textrm{eff}}^{N/2} \ket{0,N}} {\prod_{j=1}^{N/2-1}(E_0-E_{2j})} = \frac{U}{L} \cdot \left(\frac{i \eta}{2}\right)^{N/2} \cdot \frac{N!}{\prod_{j=1}^{N/2-1} (2j)^2}.
\label{gap}
\end{equation}
Here, $E_j = \frac{U}{2L} \left( 2 j (N-j) + N^2 - N\right)$ is the diagonal interaction energy shift. Finally, the term $\xi$ in Eq.~(\ref{ham2by2}) is the energy shift induced by the non-diagonal terms in the effective Hamiltonian (\ref{heff})
\begin{widetext}
\begin{equation}
\xi = -\frac{|\bra{N,0} \op{H}_{\textrm{eff}} \ket{N-2,2}|^2} {(E_0-E_{2})}+\left(\frac{|\bra{N,0} \op{H}_{\textrm{eff}} \ket{N-2,2}|^4} {(E_0-E_{2})^3}-\frac{|\bra{N,0} \op{H}_{\textrm{eff}} \ket{N-2,2}|^2 |\bra{N-2,2} \op{H}_{\textrm{eff}} \ket{N-4,4}|^2} {(E_0-E_{2})^2 (E_0-E_{4})}\right)+\dots
\end{equation}
\end{widetext}
As $\xi$ is just a constant in the $2\times 2$ subspace we do not need to evaluate it explicitly and neglect it hereafter.

The ground state of the two-by-two Hamiltonian (\ref{ham2by2}) is given by
\begin{equation}
\frac{c_0 \ket{N,0} + i^{N/2} c_N \ket{0,N}}{\sqrt{2}}
\end{equation}
where the ratio of its amplitudes is given by
\begin{equation}
\frac{c_0}{c_N} = \frac{\Delta E - \sqrt{(\Delta E)^2 + |2 \Delta|^2}}{|2\Delta|}
\end{equation}
and the energies of the ground and excited states are $E_\pm = \pm \sqrt{\Delta E^2 + |2\Delta|^2}/2$. We see that to obtain a cat-like superposition, i.e.~$c_0/c_N \approx 1$, the energy difference must not dominate over the coupling $|\Delta E| \ll |2\Delta|$. In this limit the energy gap $E_+ - E_-$ is given by $E_+ - E_- \approx |2\Delta|$.

\begin{figure}
 \centering
 \includegraphics[width=\columnwidth]{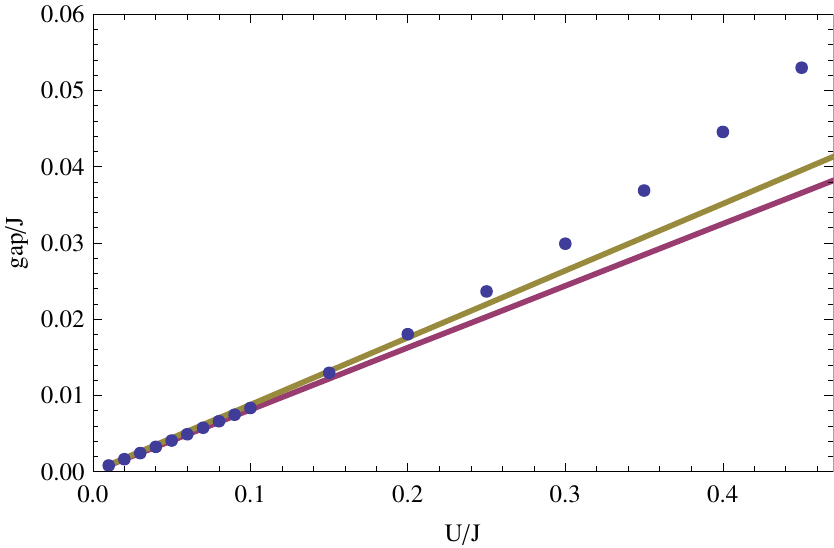}
 \caption{(Color online) Energy gap for $L=N=4$, $t/J=0.7$, $\theta = \pi/4$ as a function of $U/J$ from exact diagonalization (blue points), effective Hamiltonian (red line) and the gap formula (yellow line).}
 \label{fig:gap}
\end{figure}

In Fig.~\ref{fig:gap} we plot the energy gap between the two lowest-energy states of the many-body spectrum as a function of the on-site interaction $U/J$ for $L=N=4$, $t/J = 0.7$ and $\theta=\pi/4$. We find that the analytic formula (\ref{gap}) agrees very well with the diagonalization of the effective Hamiltonian (\ref{heff}) and at small interaction strengths $U < 0.2$ also with the exact diagonalization of the full many-body Hamiltonian (\ref{ham}).

Let us compare our result (\ref{gap}) with the coupling $\Delta$ in the case of a uniform ring lattice ($t=J$). Unfortunately, an analytic expression equivalent to Eq.~(\ref{gap}) has not been found in this case but it is known \cite{Rey07} that the coupling $\Delta$ in first non-vanishing order for unity filling $N/L = 1$ is proportional to $\Delta/J \propto (\frac{U}{2LJ})^{N-1}$. In both cases the coupling $\Delta$ decreases exponentially with the number of particles $N$. This is because multiple two-particle scattering processes are the microscopic origin of the coupling $\Delta$ and transitions between the two configurations $\ket{N,0}$ and $\ket{0,N}$ are thus highly off-resonant $N/2$-th and $N-1$-th order processes, respectively. Consequently, the energy gap vanishes in the thermodynamic limit $N \rightarrow \infty$ and cat state production both in uniform and slightly non-uniform ring lattices is restricted to modest numbers of atoms. However, since the energy gap in first non-vanishing order is proportional to $U(\eta/2)^{N/2}$ for slightly non-uniform rings as compared to $U(U/J)^{N-1}$ for uniform rings, it can be one order of magnitude bigger in the relevant parameter regime. Furthermore, we point out that the appearance of cat states in slightly non-uniform rings is not limited to commensurate filling as long as the number of particles $N$ is even.

\begin{figure}
 \centering
 \includegraphics[width=\columnwidth]{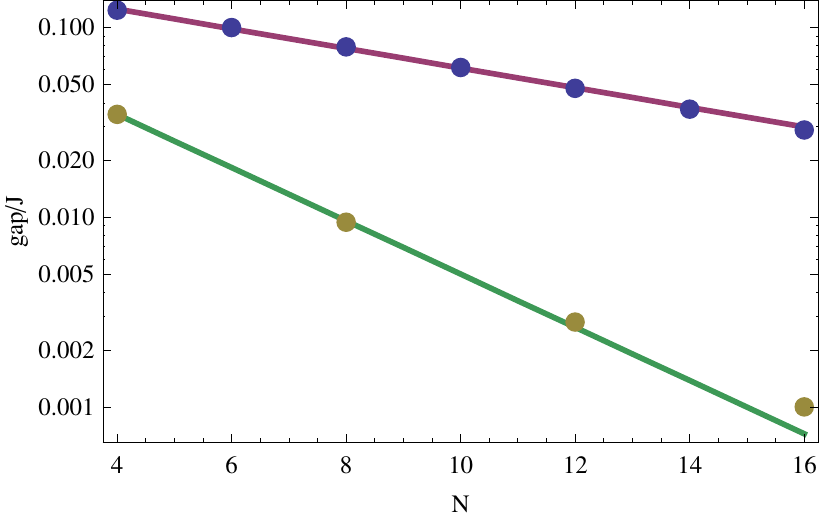}
 \caption{(Color online) Energy gap as a function of particle number $N$ with $t/J = 0.7$ (upper curve) and $t/J =1$ (lower curve) for the ring of $L=4$ sites and $U/J =0.5$. The solid lines are exponential fits to guide the eye.}
 \label{fig:gapscaling}
\end{figure}

In Fig.~\ref{fig:gapscaling} we plot the energy gap $L=4$ and $U/J = 0.5$ as a function of number of particles $N$ for $t/J=0.7$ and $t/J=1$. The solid line is an exponential fit to the points to guide the eye. We see that by changing from $t/J = 1$ to $t/J = 0.7$ the energy gap increases by over one order of magnitude for numbers of particles $N \ge 8$. As experiments are typically limited to $1\textrm{sec}$ and typical hopping energies are of order of $0.05 E_r\sim 10^{2} \mathrm{Hz}$, the detection of cat states by non-equilibrium dynamics is limited to $\Delta/J =10^{-2}$. This in turn restricts our cat-state production scheme to a few tens of particles in contrast to less than ten particles for uniform ring lattices (see Fig.~\ref{fig:gapscaling}).

\subsection{Product states in the limit $t/J = 0$}

Let us now try to understand the properties of the slightly non-uniform ring by looking at the opposite limit $t/J = 0$ where the system becomes an array of $L/2$ isolated double-well potentials. In this section we will focus on the case of $L=4$ sites for notational convenience, but our arguments are valid for all $L$ and $N$.

Since the double-wells are isolated for $t/J = 0$, we can treat the atoms in each double-well separately. Without interactions $U=0$ the states $\ket{n,N-n}$ ($0 \le n \le N$) with $n$ atoms in the ground state of the first and $N-n$ in the ground state of the second double-well have the same single-particle energy. In first-order perturbation theory the on-site interaction shifts the ground state energy of a double-well with $n$ particles by $U/2 \cdot n (n-1)/2$ and thus breaks this degeneracy. The energies of the states $\ket{n,N-n}$ become
\begin{equation}
E_{n,N-n} = \frac{U}{4} \cdot \left( N^2-N-2n(N-n) \right).
\label{dwenergy}
\end{equation}
As expected, repulsive interactions favor states of equal numbers in the two double-wells, so that the states $\ket{N/2,N/2}$ and $\ket{N/2 \pm 1,N/2 \mp 1}$ are the lowest-lying states. The energy difference between the ground and the two-fold degenerate excited state is thus $U/2$. We note that the opposite sign of the energy shift due to the on-site interaction in Eqs.~(\ref{heff}) and (\ref{dwenergy}) stem from the two different single-particle basis sets used to span the two-mode Fock states \footnote{One simple example where this effect occurs is the two-mode approximation for $N$ particles in a double-well potential. In the site basis $\ket{n_L,N-n_L}$ with $n_L$ particles on the left and $N-n_L$ particles on the right-hand side of the double-well, the energy shift due to the on-site interaction Hamiltonian (\ref{hamint}) is $\bra{n_L,N-n_L} \op{H}_{\textrm{int}} \ket{n_L,N-n_L} = U/2 \cdot (N^2 - N - 2 n_L (N-n_L))$ which shows the minus sign similar to Eq.~(\ref{dwenergy}). Changing the single-particle basis set to quasi-momentum states using $\des{b}_\pm=(\des{a}_L \pm \des{a}_R)/\sqrt{2}$ we obtain $\bra{n_+,N-n_+} \op{H}_{\textrm{int}} \ket{n_+,N-n_+} = U/4 \cdot (N^2 - N + 2 n_+ (N-n_+))$ where $n_+$ particles are in the symmetric and $N-n_+$ particles are in the anti-symmetric orbital and which shows the plus sign similar to Eq.~(\ref{heff}).}.

\begin{figure}
 \centering
 \includegraphics[width=\columnwidth]{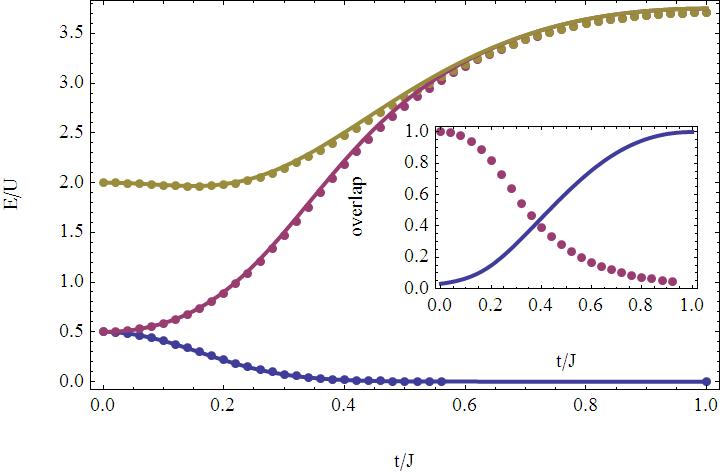}
 \caption{(Color online) Excitation energies for $L=4$, $N=16$ and $\theta = \pi/4$ as a function of $t/J$ from the effective Hamiltonian (\ref{heff}) (solid line) and the full Hamiltonian (\ref{ham}) (dotted line). The inset shows for the same parameters the overlap of the ground state with the ground state of the isolated double-wells $t/J = 0$ (red dotted line) as well as the uniform ring $t/J = 1$ (blue solid line) as a function of $t/J$.}
\label{fig:spectrumvst}
\end{figure}

In Fig.~\ref{fig:spectrumvst} we compare the excitation energies of the effective Hamiltonian (\ref{heff}) with the full many-body Hamiltonian (\ref{ham}) as a function of $t/J$ and find that they agree for all ratios $t/J$. Furthermore, we see that in the limit $t/J \rightarrow 0$ the excitation energies agree with the picture of the isolated double-wells. In the inset of Fig.~\ref{fig:spectrumvst} we plot the overlap of the ground state of the system with the ground state of the isolated double-wells $t/J = 0$ as well as the uniform ring $t/J = 1$. We find that as the ratio $t/J$ decreases, the overlap with the cat-like superposition of two quasi-momentum states decreases while the overlap with the ground state of the isolated double-wells increases.

We further note that in the limit $t/J \rightarrow 0$ the physics is independent of the number of particles $N$ and the applied phase twist $\theta$ (see Eq.~(\ref{spspectrum})). As the exact ground state of the effective Hamiltonian (\ref{heff}) for arbitrary $t/J$ and $N$ is very non-trivial, a legitimate way to gain some physical insight about its properties is to interpolate between the two limits discussed above. Using this perspective one can attribute the less severe scaling of the energy gap with the number of particles, to hybridization of the cat and product states.

\section{Bigger energy gaps and weaker sensitivity}

A fair comparison between cat state production in uniform and non-uniform ring lattices has to contrast the gain in energy gap with the degradation of the cat fidelity or "cattiness". To quantify the "cattiness" of a quantum state we consider two measures: (i) we define the cattiness of the first kind as the overlap of the ground state with the quasi-momentum cat state
\begin{equation}
\ket{\psi_1} = \frac{1}{\sqrt{2 N!}} \left[ \left(\cre{b}_0\right)^N + \left(\cre{b}_{L/4}\right)^N \right] \ket{\mathrm{vac}}
\end{equation}
and (ii) we define the cattiness of the second kind as the overlap of the ground state with the cat-like superposition of the two single-particle modes
\begin{equation}
\ket{\psi_2} = \frac{1}{\sqrt{2 N!}} \left[ \left(\cre{c}_0\right)^N + i^{N/2} \left(\cre{c}_{L/4}\right)^N \right] \ket{\mathrm{vac}}
\end{equation}
which depends on the ratio $t/J$ as defined in Eq.~(\ref{definec}).

In this section we demonstrate that while there is a trade-off between bigger energy gaps and cattiness of the first kind, non-uniform ring lattices offer bigger energy gaps at fixed cattiness of the second kind. Consequently, they are less sensitive to detunings of the rotation frequency which is an important practical advantage. On the other hand it is clear that to detect a cat state in two single-particle modes which do not have well defined quasi-momentum is more demanding experimentally.

Both cat-state measures can be calculated perturbatively in the limit of weakly non-uniform rings $t/J \approx 1$ and weak on-site interactions $U/J \ll 1$. The quasi-momentum cat state $\ket{\psi_1}$ is the ground state for $t/J = 1$ and $U/J \rightarrow 0$. Deviations are due to the off-diagonal part of the single-particle Hamiltonian in quasi-momentum representation (\ref{hamsp}) as well as the on-site interactions (\ref{hamint}). We obtain for the overlap
\begin{equation}
\left| \braket{\psi_1}{\psi_0} \right|^2
= 1- \frac{N}{4} \left( \frac{J-t}{J+t} \right)^2 - \left( \frac{U\sqrt{N (N-1)}}{2 L \sqrt{J^2 + t^2}} \right)^2
\label{depletion1}
\end{equation}
where $\ket{\psi_0}$ denotes the ground state of the system at finite $t/J$ and $U/J$. In case of the cattiness of the second kind, depletion is due to the off-diagonal terms of the effective Hamiltonian (\ref{heff}) as well as the part of the interaction Hamiltonian (\ref{hamint}) which couples states within the subspace to states outside the subspace of the effective Hamiltonian. We obtain
\begin{equation}
\left| \braket{\psi_2}{\psi_0} \right|^2 = 1 - \frac{N(N-1) \eta^2}{8(N-2)^2} - \left( \frac{U\sqrt{N (N-1)}}{2 L \sqrt{J^2 + t^2}} \right)^2.
\label{depletion2}
\end{equation}

\begin{figure}
 \centering
 \includegraphics[width=\columnwidth]{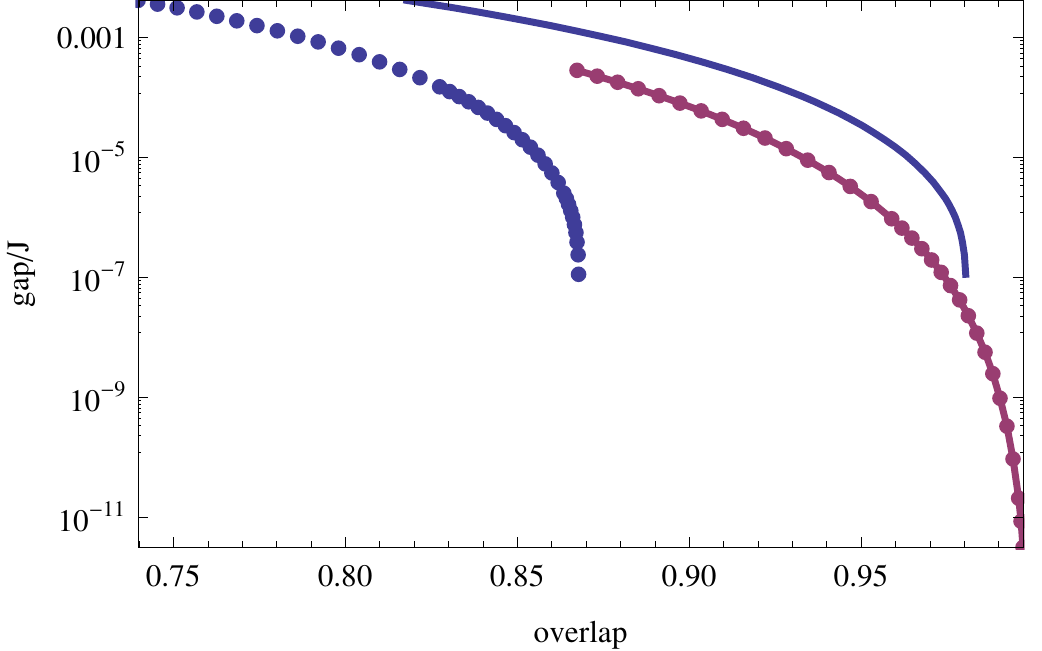}
 \caption{(Color online) Energy gap versus cat-state measure of the first kind (dotted points) as well as of the second kind (solid lines) for $t/J = 0.7$ (upper blue) and $t/J = 1$ (lower red). The other parameters are $L=4$, $N=16$ and $\theta = \pi/4$.}
 \label{fig:depletionvsgap}
\end{figure}

In Fig.~\ref{fig:depletionvsgap} we plot the energy gap versus the two cat-state measures for $t/J = 0.7$ and $t/J = 1$, respectively. The other parameters are $L=4$, $N=16$, $\theta = \pi/4$. We find that: (i) the larger $U/J$, the bigger the energy gap but the smaller are both cat-state measures. This is readily understood from Eqs.~(\ref{gap}), (\ref{depletion1}) and (\ref{depletion2}): stronger on-site interactions increase the coupling $|\Delta|$ and, therefore, increase the energy gap. However they also increase coupling to other states outside of the subspace of $\op{H}_{2\times 2}$ and the cat state is gradually depleted \cite{Hallwood06b}. (ii) The smaller $t/J$, the bigger the energy gap but the smaller is the cattiness of the first kind at fixed $U/J$. This is in agreement with Eqs.~(\ref{gap}) and (\ref{depletion2}). Physically, this means that the off-diagonal part of the single-particle Hamiltonian increases the coupling $|\Delta|$ but also dilutes the many-body correlation in the quasi-momentum basis. Consequentially, there is a trade-off between  energy gap and cattiness of the first kind. We note that it might be a sizeable advantage to sacrifice part of the cat-like correlation (and therefore a weaker signal in the many-body oscillations we discuss in the next section) but make it visible on experimentally accessible time scales. Finally, we find that (iii) one can increase the energy gap by about an order of magnitude (for $N=16$ particles) at fixed cattiness of the second kind when the ratio $t/J$ is changed from $t/J = 1$ to $t/J = 0.7$. While we believe that this is the natural measure of comparison between uniform and non-uniform ring lattices, we admit that probing the cat state in the single-particle basis is a non-trivial task, since -- as we will show below -- time-of-flight imaging maps more closely on the quasi-momentum basis.

\begin{figure}
 \centering
 \includegraphics[width=\columnwidth]{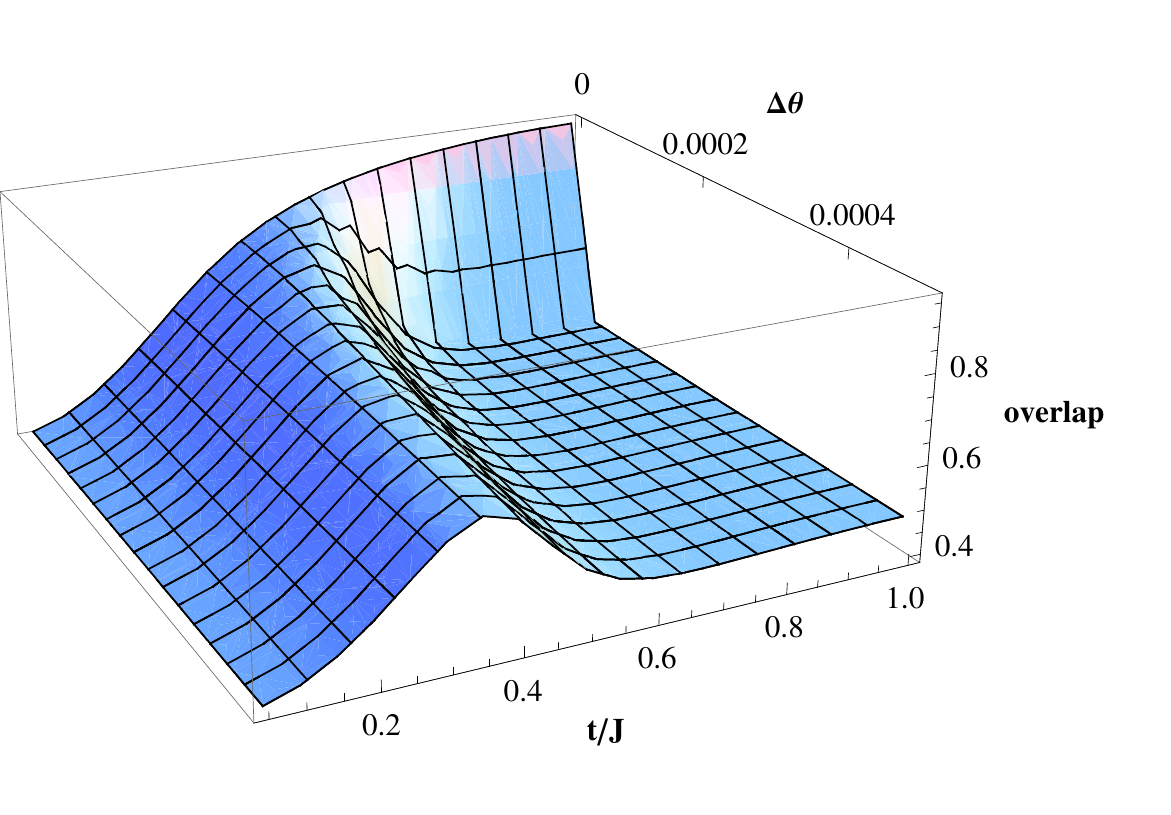}
 \caption{(Color online) Cat-state measure of the second kind as a function of $t/J$ and the detuning $\Delta \theta$ while $L=4$, $N=16$ and $U = 0.2$ are fixed.}
 \label{fig:offres}
\end{figure}

Increasing the coupling $|\Delta|$ with non-uniform ring lattices has important implications for the sensitivity to detunings of the rotation frequency away from the critical phase twist. In Fig.~\ref{fig:offres} we plot the cat-state measure of the second kind as a function of $t/J$ and the detuning $\Delta \theta$ while $L=4$, $N=16$ and $U = 0.2$ are fixed. We find that the system is much less sensitive to detunings away from the critical phase twist for $t/J < 1$, i.e.~the requirements on the phase control precision $\Delta \theta$ are relaxed. To understand how this comes about let us go back to the two-by-two Hamiltonian (\ref{ham2by2}). Similar to the two-state Hamiltonian for a single particle in a double-well potential, its ground state depends on the ratio of energy difference $\Delta E$ to the coupling energy $|\Delta|$: for $|\Delta E| \gg |\Delta|$ the system is in one of the states $\ket{N,0}$ (in the double-well analogy on the left-hand side) or $\ket{0,N}$ (on the right hand side) whereas it is in their superposition state in the limit where the coupling energy dominates $|\Delta E| \ll |\Delta|$. As the coupling $|\Delta|$ in Eq.~(\ref{gap}) decreases exponentially with increasing number of particles $N$, more control over the rotation frequency is needed to tune systems with more particles into resonance, i.e.~to make the energy difference $|\Delta E|$ between the states $\ket{N,0}$ and $\ket{0,N}$ smaller than their coupling $|\Delta|$. This was identified as one of the barriers for creating large cat-like states \cite{Hallwood06b}. As the gap increases when $t/J$ is decreased (see Eq.~(\ref{gap})), we conclude that using non-uniform ring lattices $t \not = J$ can improve the situation on this issue.

\section{Dynamical detection of cat-like correlations}\label{secdy}

In this section we propose to induce many-body oscillations by suddenly changing the applied phase twist $\theta$ in order to detect the coherent superposition of two quasi-momentum states $\ket{N,0} = ( \cre{b}_0 )^N \ket{\textrm{vac}} / \sqrt{N!}$ and $\ket{0,N} = ( \cre{b}_{L/4} )^N \ket{\textrm{vac}} / \sqrt{N!}$ at the anti-crossing of the many-body spectrum.

If we assume the system is initially in the ground state of the full Hamiltonian at $\theta = 0$ which will predominantly be the state $\ket{N,0}$, i.e.~$\ket{\psi(t=0)} \approx \ket{N,0}$. Then, the phase twist is changed to $\theta = \pi/4$, where the eigenstates are approximately $\ket{\pm} \approx (\ket{N,0} \pm \ket{0,N})/\sqrt{2}$. The time evolution of the probability to be in the state $\ket{N,0}$ and $\ket{0,N}$ is
\begin{equation}
|b_0(t)|^2 = \left|\braket{N,0}{\psi(t)}\right|^2 \approx \frac{1 + \cos \nu t}{2}
\end{equation}
and $|b_N(t)|^2 \approx 1 - |b_0(t)|^2$ where $\hbar\nu = E_+ - E_-$ denotes the energy gap and where the approximate signs indicate that we neglect the depletion due to the non-diagonal part of the single-particle Hamiltonian (\ref{hamsp}) as well as the on-site interaction Hamiltonian (\ref{hamint}).

\begin{figure}
 \centering
 \includegraphics[width=\columnwidth]{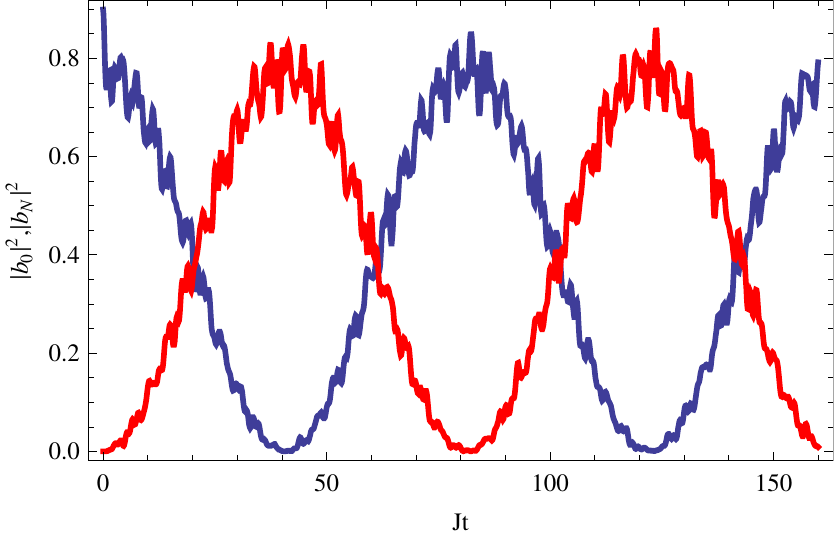}
 \caption{(Color online) Many-body dynamics $|b_0(t)|^2$ and $|b_N(t)|^2$ for $L=4$ and $N=8$ following a sudden change in phase twist from $\theta = 0$ to $\theta=\pi/4$ at $U/J =1$ and $t/J = 0.7$.}
 \label{fig:dynamics}
\end{figure}

In Fig.~\ref{fig:dynamics} we show the many-body dynamics, i.e.~$|b_0(t)|^2$ and $|b_N(t)|^2$ for $L=4$ and $N=8$ following a sudden change in phase twist from $\theta = 0$ to $\theta=\pi/4$ at time $t=0$ with $U/J =1$ and $t/J = 0.7$. We see that the many-body oscillations following this sudden change occur at the frequency of the gap $\hbar \nu$. These oscillations are modulated by an oscillation with frequency $U (N-1)/L$ which is the energy difference between the nearly-degenerate ground states and the next lowest-lying intermediate states $\ket{2,N-2}$ and $\ket{N-2,2}$. The amplitude of the oscillations is smaller than one, i.e. $|b_0(t)|^2 + |b_N(t)|^2 < 1$, due to the depletion of the quasi-momentum cat state $|\braket{\psi_1}{\psi_0}|^2$ (see Eq.~(\ref{depletion1})).

\section{Detection}

In this section we study the spatial density profile of an atomic cloud when it is let to freely expand after turning off the ring lattice potential. Our analysis shows that states with different quasi-momentum states exhibit a distinctive time-of-flight pattern and that therefore the latter can be used as an experimental probe for detecting the many-body oscillations between the $|0,N\rangle$ and  $|N,0\rangle$ discussed in previous section.

For this discussion we will consider the expansion of the wavefunction of a single atom initially confined in a 1D lattice ring geometry. For simplicity we approximate the axial, radial and angular Wannier functions as Gaussians with width $\sigma_z$, $\sigma_r$ and $\rho_0 \sigma_\Theta$, respectively, and localized in a ring of radius $\rho_0$ at $z=0$
\begin{equation}
\psi({\bf{r}},t=0) = A e^{-\frac{z^2}{2\sigma_z^2}} e^{-\frac{(r-\rho_0)^2}{2\sigma_r^2}} \sum_{n=1}^L c_n e^{-\frac{(\Theta-2\pi n/L)^2}{2\sigma_\Theta^2}}
\end{equation}
where $A$ is a normalization constant. If at time $t=0$ the atom is in an eigenstate of the uniform ring with quasi-momentum $\hbar 2\pi q_0/L a $ then $c_n = \frac{1} {\sqrt{L}} e^{i (2 \pi n q_0/L)}$.

At sufficiently large times $t\gg M \rho_0/\hbar$ one can take the far-field limit and approximate the density of the expanded cloud as being proportional to the momentum distribution of the initial state, i.e.
\begin{equation}
|\psi({\bf{r}},t \to \infty )|^2\approx \left (\frac{M}{ht} \right)^3|\phi({\bf{Q(r)}})|^2
\end{equation}
where ${\bf{Q(r)}}=M{\bf{r}}/\hbar t$ and $|\phi({\bf{k}})|^2$ is the momentum distribution of the initial state. After integrating along the $z$ axis, one can write a simple analytical expression of the transverse spatial density distribution in the limit of strongly confined Wannier orbitals (i.e.~$\sigma_r \to 0$ and  $\sigma_\Theta \to 0$)
\begin{eqnarray}
\lefteqn{ \left| \psi(\rho,\Theta, t \to \infty ) \right|^2
= \int|\psi({\bf{r}},t \to \infty )|^2 dz} \nonumber \\
& & \propto \left (\frac{M \rho_0}{Lht} \right)^2\left |\sum_n e^{ (i \frac{2 \pi n q_0}{L}-i Q(\rho) \rho_0 \cos( \frac{2 \pi n q_0}{L}-\Theta))}\right |^2.
\label{expo}
\end{eqnarray}

In the limit of a large number of lattice sites $L$, the sum in Eq.~(\ref{expo}) can be approximated by an integral and the time-of-flight profile reduces to
\begin{eqnarray}
|\psi(\rho,\Theta, t \to \infty )|^2 \to \left (\frac{M \rho_0}{ht} \right)^2\left |J_{q_0}[Q( \rho)\rho_0] \right|^2
\end{eqnarray}
where $J_{q_0}(x)$ are Bessel functions of the first kind. This case corresponds to the rotationally symmetric case discussed in Ref.~\cite{Cozzini06} where the position of the zeros and maxima of the Bessel functions provide a full characterization of the density profile. A state initially with zero quasi-momentum for example will exhibit an interference peak at the origin while a state with non-zero quasi-momentum will exhibit a central hole. As the position of the first maximum of $J_{q_0}[Q(\rho) \rho_0]$ is an increasing function of $q_0$, the larger the initial quasi-momentum the wider the central hole. For a finite number of lattice sites the sum does not correspond exactly to a Bessel function and the momentum distribution does not become fully radially symmetric. However, there is still a unique correspondence between the position of the peaks in the absorption image and the initial quasi-momentum distribution (see also Ref.~\cite{Peden07}). Consequently, the latter provides a means to experimentally determine the quasi-momentum of the wavefunction before the release.

\begin{figure*}[ht]
\begin{center}
\leavevmode {\includegraphics[width=4.5 in]{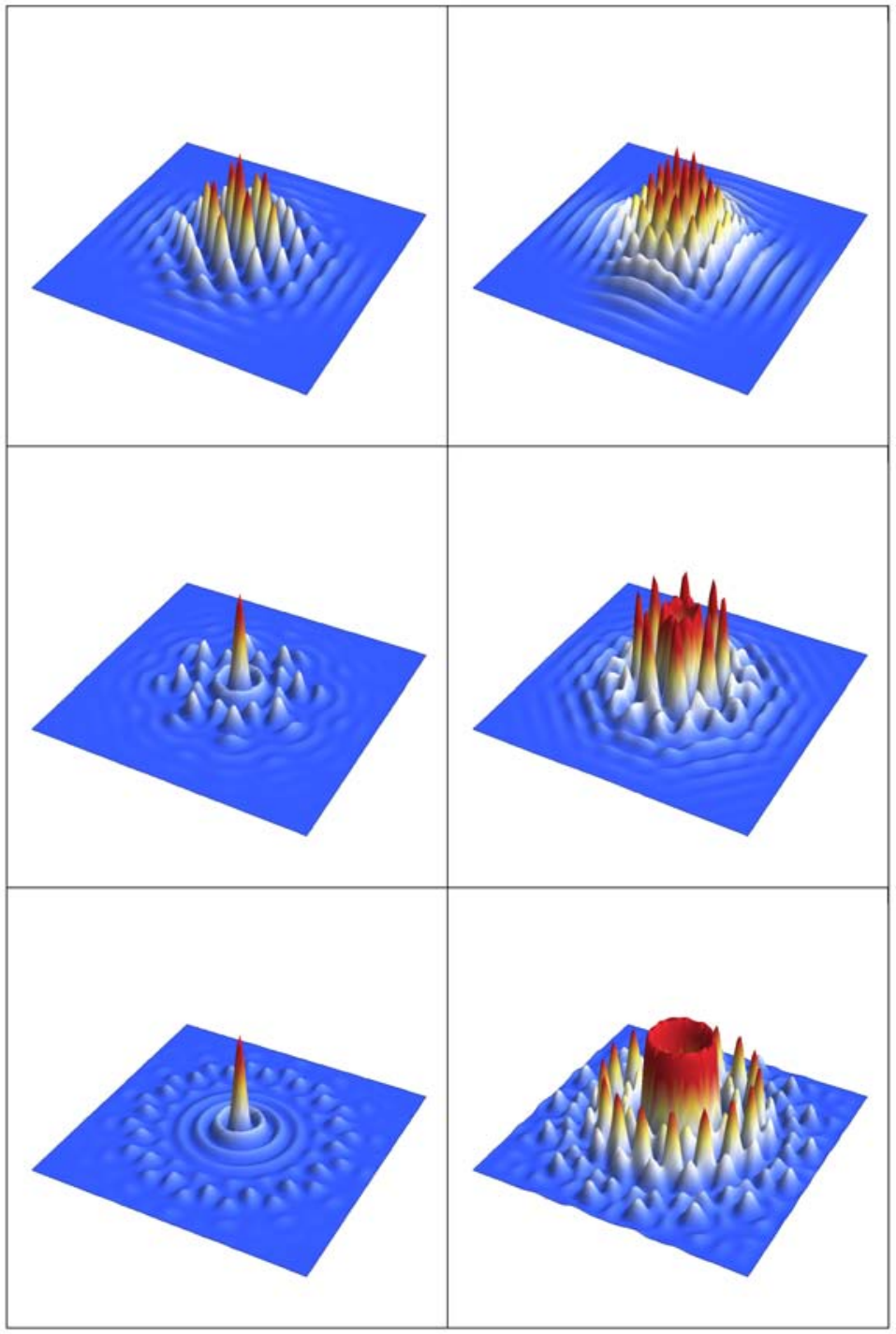}}
\end{center}
\caption{(Color online) Transverse time-of-flight images of an initial state with quasi-momentum $q_0=0$ (left) and $q_0=L/4$ (right) for different ring lattice geometries (top $L=4$, center $L=8$ and bottom $L=16$).}
\label{fig:expansion}
\end{figure*}

In Fig.~\ref{fig:expansion} we show the numerically calculated time-of-flight images of different ring lattice geometries for two states with initial quasi-momentum $q_0 = 0$ and $L/4$, respectively. These states correspond to the two types of initial distributions that one has to experimentally distinguish in order to probe the many-body oscillations we have discussed in the previous section. The figure shows the distinctive interference pattern of each of the two initial states and the asymptotic approach to radial symmetry as the number of lattice sites $L$ is increased.

\section{Conclusion}

We have investigated cat state production with ultracold bosons in rotating ring superlattices and found the improvement compared to uniform ring lattices to be threefold: the energy gap between the cat-like ground and first excited states scales more favorably with the number of particles and the constraints on phase twist control and commensurate filling are relaxed. Finally, we have shown that the different quasi-momentum states can be distinguished in time-of-flight absorption images and proposed to probe the cat-like correlations via the many-body oscillations induced by a sudden change in the rotation frequency. Since the exponential scaling of the energy gap with the number of particles remains, we suspect that cat state production in ultracold atomic gases is limited to modest number of atoms at least for systems with contact interactions. One possible way-out is to consider systems with long-range interactions such as polar molecules with dipolar interactions.

\begin{acknowledgments}
We thank Liang Jiang, Mohammad Hafezi, Mikhail Lukin and Steven Girvin for useful comments and discussions. This work was partially supported by the National Science Foundation through a grant for the Institute for Theoretical Atomic, Molecular and Optical Physics (ITAMP) at Harvard University and Smithsonian Astrophysical Observatory. A.N.~is grateful for the hospitality of the ITAMP visitor's program. A.N.~acknowledges a scholarship from the Rhodes Trust, A.M.R.~an ITAMP fellowship, and K.B.~support from the Royal Society and the Wolfson Foundation.
\end{acknowledgments}


\end{document}